\documentclass[aps,prl,twocolumn,showpacs,floatfix,preprintnumbers,nofootinbib,superscriptaddress]{revtex4}

\usepackage{graphicx}
\usepackage{color}
\usepackage{natbib}
\usepackage{multirow}

\begin{document}

\preprint{CERN-PH-TH/2011-236}

\title{Apparent faster than light propagation from light sterile neutrinos}

\author{Steen Hannestad}
\affiliation{Department of Physics and Astronomy,
 University of Aarhus, 8000 Aarhus C, Denmark}

\author{Martin S. Sloth}
\affiliation{
D\'{e}partment de Physique Th\'{e}orique and Center for Astroparticle Physics,\\
 Universit\'{e} de Gen\`{e}ve, 24 Quai E. Ansermet, CH-1211 Gen\`{e}ve, Switzerland}

\affiliation{
CERN, Physics Department, Theory Unit, CH-1211 Geneva 23, Switzerland}
\date{September 28, 2011}

\begin{abstract}
Recent data from the OPERA experiment seem to point to neutrinos propagating faster than light. One possible physics explanation for such a result is the existence of light sterile neutrinos which can propagate in a higher dimensional bulk and achieve apparent superluminal velocities when measured by an observer confined to the 4D brane of the standard model. Such a model has the advantage of easily being able to explain the non-observation of superluminal neutrinos from SN1987A. Here we discuss the phenomenological implications of such a model and show that it can provide an explanation for the observed faster than light propagation of neutrinos.
\end{abstract}

\maketitle

\section{Introduction}

Recent data from the OPERA experiment seem to point to neutrinos propagating faster than light, with $|v-c|/c \sim
2.5 \times 10^{-5}$ \cite{:2011zb}. This result is in concordance with the tentative detection of superluminal neutrino propagation by the MINOS experiment \cite{Adamson:2007zzb}. While this is seemingly in conflict with special relativity in 4 dimensions, a number of possible explanations for Lorentz violation exist in the literature
\cite{Ellis:2003if,Kostelecky:1988zi,Chung:2000ji,Csaki:2000dm,Horava:2009uw,Giudice:2010zb}.

Here, we propose that the faster than light propagation of muon neutrinos is caused by mixing with light sterile neutrinos.
In models with large extra dimensions, standard model particles are confined to a brane with 3 spatial dimensions because of gauge charges. This is true also for the standard active neutrinos. However, a sterile neutrino is a gauge singlet in the standard model and can in principle freely propagate in the bulk. A number of models exist where bulk geodesics beginning and ending on the brane are shorter than the corresponding 4D geodesics traversed by particles on the brane and where bulk particles can therefore travel superluminally.

In the next section, discuss heavy vs light sterile neutrinos in the context of the new experimental results. In Section III we describe one particular brane world model which allows for superluminal motion of bulk particles. In Section IV we discuss the oscillation phenomenology in a simple 2-neutrino mixing scheme, involving only $\nu_\mu$ and a new sterile component. Finally, Section V contains our conclusions.

\section{Light sterile neutrinos}

Since the constraints on Lorentz violation involving photons and electrons are very strong, we are searching for an effect of Lorentz violation effectively restricted to neutrinos. One of the simplest ways to achieve this goal, it is to introduce a sterile neutrino, and confine the Lorentz violation in this sector.

The models proposed so far to explain the apparent superluminal signal propagation in MINOS and OPERA,  has focussed on a heavy sterile neutrino with Lorentz-violating interactions and mixings with the active neutrinos suppressed by mass of the heavy sterile neutrino. The most serious drawback of these models, it is that Lorentz invariance needs to be flavor-independent in order not to destroy the observed coherent oscillations between active neutrinos. In models with a heavy sterile neutrino with mass $M$ much larger than the energy of the neutrinos, $E_{\nu}$, the heavy sterile neutrino can be integrated out and affects all propagating active neutrinos which it mixes with. If the the effect on different neutrino spices is different it will destroy the coherent oscillations. On the other hand if it is flavor independent, then one encounters severe problems with loop induced Lorentz violations for electrons, from a loop with a $W$ boson and an electron neutrino, which contradicts the very tight bounds on the speed of propagation on electrons \cite{Giudice:2011mm}.

Instead, one might therefore want to consider a model with a light sterile neutrino, which have the merit of providing a possible explanation of observed short baseline neutrino anomalies (see e.g.\ \cite{Kopp:2011qd}), as well as a possible explanation for the observed excess relativistic energy density in cosmology (see e.g.\ \cite{hamann}). Unlike heavy steriles, a light sterile neutrino state cannot be integrated out. In the case of small sterile-active neutrino mixing, only a fraction of the observed neutrinos will have propagated as sterile neutrinos. In this case the Lorentz violation need not to be flavor independent, and one might alleviate the problems with loop induced Lorentz violation in electrons, by making the sterile neutrino interacting only with muon neutrinos. This is a prediction which is testable by experiments such as MINOS or OPERA. 
However, even if Lorentz violation is restricted to the muon sector there are still potential problems with known limits on the muon velocity which is limited to be within approximately $10^{-11}$ of $c$ \cite{Altschul:2006uw}. 
The loop induced contribution to the muon velocity is expected to be of order $10^{-9}$ \cite{Giudice:2011mm}, and even though the disagreement is far better than for the electron sector where the experimental bound is of order $10^{-15}$ it is potentially problematic.

We note that the possibility of faster than light propagation of bulk sterile neutrinos was first discussed in \cite{Dent:2007rk,Pas:2006si,Pas:2005rb}, albeit in a slightly different context. In these papers the idea was to induce a large effective mixing angle to the sterile sector in order to provide an additional MSW-like resonance with which to explain short baseline disappearance experiments.

In the same way, we will see that the apparently very stringent bound from SN1987A \cite{Hirata:1987hu} is avoided. Since only a small fraction of the neutrinos actually travels as a sterile neutrino and therefore appears superluminal, it will not be observable from SN1987A. This means that the Lorentz violation does not need to have a strong energy dependence, as otherwise needed in other models in order to avoid the SN1987A constraint \cite{Giudice:2011mm}.

\section{Sterile neutrino propagation}

The details for the Lorentz breaking in the sterile neutrino sector is not so crucial to the present scenario, and in fact, contrary to other recent models, the model we are presenting is relatively insensitive to e.g. the energy dependence of the Lorentz violating terms, as mentioned above.

But in order to have a more specific toy model, one may consider a general brane-world model, with metric

\begin{equation}
ds^2 = -h(z)dt^2+g(z)d{\vec{x}}^2+f(z)dz^2~,
\end{equation}

where $(t,x^i)$ are the 4 dimensions on the brane, and $z$ is the transverse extra-dimension. It is assumed that the Standard Model of particle physics is confined on a brane while gravity propagates in the bulk, similarly to what happens in Randall-Sundrum scenarions \cite{Randall:1999ee,Randall:1999vf}.  But while those models has $h(z)=g(z)$ and an $SO(3,1)$ symmetry, such that light propagates at the same speed in the planes parallel to the brane all along the extra-dimension, in models where this symmetry is absent, the speed of light in the planes parallel to the brane varies along the extra-dimension \cite{Chung:2000ji}. In such models gravity waves can take short cuts through the extra dimensions, and can appear to propagate faster than the speed of light from an effective four dimensional viewpoint. This was explored in models of so called ``asymmetrically warped extra dimensions", with the $SO(3,1)$ symmetry broken by a black hole in the bulk \cite{Csaki:2000dm}.

We will now further consider a scenario with sterile neutrinos, which does not carry gauge charges and therefore, like gravity, does not need to be confined to the brane unlike all other standard model fermions. Intuitively, one can imagine that sterile neutrino geodesics are allowed to propagate in the bulk on geodesics, which can be shorter than the corresponding 4 dimensional geodesics traversed by photons and active neutrinos. In this case the sterile neutrino component would transverse a smaller total distance and arrive earlier, but the velocity of neutrinos still obeys $v/c(z) < 1$ in the $4+n$ dimensional bulk. However, in practice the extra dimension is typically small compared to the relevant wavelengths of sterile neutrinos, and the sterile neutrino wave function will be smeared over the extra dimension. In the effective four dimensional description, Lorentz violation will appear as a modification of the effective propagation speed of the sterile netrinos, similarly to how the effective propagation of gravitons is modified.

In the minimal version of the scenario, one would expect the effective four dimensional propagation speed of the sterile neutrino to be the same of gravitational waves, and related just to the geometry of the bulk. In the model of \cite{Csaki:2000dm}, this is related to the brane location, the charge of the black hole in the bulk, and the cosmological constant in the bulk. In the toy model of \cite{Chung:2000ji}, with $h(z)=f(z)=1$ and $g(z)=\exp(-2 b z)$, the correction to the speed of gravitational waves compared to that of light is $\delta c =bL/2$ where $L$ is the size of the extra dimension and $b$ is the warp factor.

In order to explain the MINOS and the OPERA result, one would need $\delta c > 10^{-5}$ for the sterile neutrinos, and, as mentioned, in the minimal version of this scenario, one would expect that gravitational waves experience the same effect. Therefore it is interesting to consider constraints on the propagation of gravitational waves. Some of the strongest constraints comes from direct bounds from solar-system and binary-pulsar test, which requires $\delta c < 10^{-6}$. The stronger constraints comes from requiring that ecliptic and solar equatorial planes do not precess relative to each other throughout the history of the solar system \cite{Nordtvedt}(for a review, see \cite{Will:2001mx}). But if one insist only on terrestrial bounds, or bounds from the gravitational radiation of binary pulsars, a weaker bound applies $\delta c < 10^{-3}$. Finally one needs also to consider more model dependent constraints from gravitons in the loops of Standard Model particles \cite{Burgess:2002tb}.

In a non-minimal model, one might relax some of these constraints by having non-gravitational charges for the sterile neutrino in the bulk.

\section{Oscillation phenomenology}

\subsection{Production and detection of sterile states}

{\it Production ---} In all current long baseline experiments such as K2K \cite{k2k}, MINOS \cite{minos} and OPERA \cite{opera}, neutrinos are produced from protons in a beam dump leading to pion decay and the eventual production of $\nu_\mu$ and $\nu_e$. The charged current production vertex for muon neutrinos contains a mixture of mass states, with the mainly sterile $m_4$ component entering with a relative rate of $\sin^2 \theta_{\rm eff}$ (i.e.\ the vertex amplitude is proportional to $\sin \theta_{\rm eff}$.

{\it Detection ---} The detector again uses charged current absorption (and subsequent charged lepton production) to measure the arrival of a neutrino. Neutrinos propagating in the luperluminal mass state $m_4$ have a finite probability for undergoing a charged current interaction, producing a final state muon. The relative rate is again given by $\sin^2 \theta_{\rm eff}$.

{\it Propagation ---} After the initial production, the wavepackets corresponding to different mass states, propagate with different velocities, one of them superluminally. In the simplest 2-neutrino mixing scheme where we have only $\nu_\nu$ and $\nu_s$ the oscillation length is approximately $l_{\rm osc} \sim E_\nu/\delta m^2$. For a baseline of 730 km, and energy $E_\nu \sim 30$ GeV any mass difference $\delta m^2 \gtrsim 0.1$ eV$^2$ corresponds to many oscillations along the path from emission to detection and one should expect approximate decoherence of the $m_4$ wavepacket from the other states.
Current short baseline disappearance experiments show tentative evidence for the presence of sterile neutrinos in the eV range. It would perhaps be natural to look to the same source for an explanation of the apparent superluminal signal propagation seen in MINOS and OPERA.

In this case, the sterile component contributes approximately $\sin^4 \theta_{\rm eff}$ of the total number of events detected as muon neutrinos. Interestingly, the fraction of neutrinos measured as having superluminal velocity is energy independent in the limit of $l \gg l_{\rm osc}$. This prediction is very different from other proposed explanations of the superluminal propagation.

\subsection{Constraints on $\theta_{\rm eff}$}

Current constraints on the mixing of low mass sterile neutrinos are relatively weak, with the effective $4\mu$ mixing element being constrained to be less than approximately $\sin ^2 \theta \lesssim 0.1-0.2$ (see e.g.\ \cite{Giunti:2011hn}). With such a relatively small mixing angle the SN1987A bound on neutrino velocity does not apply because only at most a fraction $\sin ^4 \theta_{e4}$ would travel in mass state 4, but be observed as $\bar \nu_e$. This would in practise mean less than 1 event observed at some earlier epoch before the main supernova event which in practise would never pose a problem. First of all it would not dilute the actual signal (which in any case could be lessened by a significant amount without being in serious conflict with current supernova models), second a small number of events occurring much earlier would not be attributable to the supernova event itself since the detection did not contain any directional information.

\section{How large a fraction of the neutrinos must travel superluminally?}

While the analysis has been carried out using the assumption that all neutrinos travel at the same velocity one could in principle perform the same analysis, but with two different neutrino components.
To achieve the same average value one would have to have the superluminal component travel at a somewhat higher velocity
\begin{equation}
\delta v_{\rm obs} = \delta v_{\rm s} n_{\rm s}/N \sim \delta v_{\rm s} \sin^4 \theta,
\end{equation}
such that for example for $\sin^2 \theta = 0.2$ one would find $\delta v_{\rm s} \sim 6 \times 10^{-4}$ which is potentially problematic. However, a specific, quantitative analysis remains to be carried out.
This indeed is probably the main caveat in this model: Too small a mixing angle makes the effect unobservable whereas too large a mixing angle would lead to disagreement with muon neutrino disappearance bounds.

\section{Discussion}

While the validity of the claim that neutrinos can travel faster than light remains to be verified by other experiments, we have shown that in the presence of sterile neutrinos and extra dimensions the result is not necessarily at odds with special relativity, and does not necessarily break Lorentz invariance at the fundamental level in the higher dimensional bulk (although it does break Lorentz invariance on the brane).

Unlike models with Lorentz violation in a heavy sterile sector, the model presented here does not necessarily lead to explicit and observable Lorentz violation in all flavours. This also means that no strong energy dependence of the effect is needed to circumvent the SN1987A bound on neutrino propagation velocity, and in our model the generic prediction is indeed that the effect should be close to energy independent (in concordance with the MINOS and OPERA observations).

It remains to be tested whether our model which predicts that only a fraction of the observed muon neutrinos have propagated faster than light is compatible with a detailed analysis of the OPERA data.
It should also be noted that the model presented here has another potential problem which should be addressed in a more detailed analysis: The superluminal sterile neutrinos induce Lorentz violation in the muon at a level which is potentially in conflict with existing experimental data. 
Nevertheless, in spite of these potential problems, models with light sterile bulk neutrinos certainly seem to be good candidates for an explanation of the extremely puzzling observation of apparent faster than light neutrino propagation.



\begin{thebibliography}{99}


\bibitem{:2011zb}
  T.~Adam {\it et al.} [ OPERA Collaboration ],
  [arXiv:1109.4897 [hep-ex]].

\bibitem{Adamson:2007zzb}
  P.~Adamson {\it et al.} [ MINOS Collaboration ],
  Phys.\ Rev.\  {\bf D76}, 072005 (2007).
  [arXiv:0706.0437 [hep-ex]].







\bibitem{Ellis:2003if}
  J.~R.~Ellis, N.~E.~Mavromatos, D.~V.~Nanopoulos, A.~S.~Sakharov,
  Int.\ J.\ Mod.\ Phys.\  {\bf A19}, 4413-4430 (2004).
  [gr-qc/0312044].

\bibitem{Kostelecky:1988zi}
  V.~A.~Kostelecky, S.~Samuel,
  Phys.\ Rev.\  {\bf D39}, 683 (1989).

\bibitem{Chung:2000ji}
  D.~J.~H.~Chung, E.~W.~Kolb, A.~Riotto,
  Phys.\ Rev.\  {\bf D65}, 083516 (2002).
  [hep-ph/0008126].

\bibitem{Csaki:2000dm}
  C.~Csaki, J.~Erlich, C.~Grojean,
  Nucl.\ Phys.\  {\bf B604}, 312-342 (2001).
  [hep-th/0012143].

\bibitem{Horava:2009uw}
  P.~Horava,
  Phys.\ Rev.\  {\bf D79}, 084008 (2009).
  [arXiv:0901.3775 [hep-th]].


\bibitem{Giudice:2010zb}
  G.~F.~Giudice, M.~Raidal, A.~Strumia,
  Phys.\ Lett.\  {\bf B690}, 272-279 (2010).
  [arXiv:1003.2364 [hep-ph]].



\bibitem{Giudice:2011mm}
  G.~F.~Giudice, S.~Sibiryakov, A.~Strumia,

  [arXiv:1109.5682 [hep-ph]].

\bibitem{Kopp:2011qd}
  J.~Kopp, M.~Maltoni and T.~Schwetz,
  arXiv:1103.4570.

\bibitem{hamann}
  J.~Hamann, S.~Hannestad, G.~G.~Raffelt, I.~Tamborra, Y.~Y.~Y.~Wong,
  Phys.\ Rev.\ Lett.\  {\bf 105}, 181301 (2010).
  [arXiv:1006.5276 [hep-ph]].


%
\bibitem{Dent:2007rk}
  J.~Dent, H.~Pas, S.~Pakvasa, T.~J.~Weiler,
  [arXiv:0710.2524 [hep-ph]].

\bibitem{Pas:2006si}
  H.~Pas, S.~Pakvasa, J.~Dent, T.~J.~Weiler,
  Phys.\ Rev.\  {\bf D80}, 044008 (2009).
  [gr-qc/0603045].

\bibitem{Pas:2005rb}
  H.~Pas, S.~Pakvasa, T.~J.~Weiler,
  Phys.\ Rev.\  {\bf D72}, 095017 (2005).
  [hep-ph/0504096].

\bibitem{Hirata:1987hu}
  K.~Hirata {\it et al.} [ KAMIOKANDE-II Collaboration ],
  Phys.\ Rev.\ Lett.\  {\bf 58}, 1490-1493 (1987).

\bibitem{Randall:1999ee}
  L.~Randall, R.~Sundrum,
  Phys.\ Rev.\ Lett.\  {\bf 83}, 3370-3373 (1999).
  [hep-ph/9905221].

\bibitem{Randall:1999vf}
  L.~Randall, R.~Sundrum,
  Phys.\ Rev.\ Lett.\  {\bf 83}, 4690-4693 (1999).
  [hep-th/9906064].

\bibitem{Nordtvedt}
  K.~Nordtvedt,
  Astrophys.\ J.\  {\bf 320}, 871 (1987).
  [gr-qc/0103036].

\bibitem{Will:2001mx}
  C.~M.~Will,
  Living Rev.\ Rel.\  {\bf 4}, 4 (2001).
  [gr-qc/0103036].


\bibitem{Burgess:2002tb}
  C.~P.~Burgess, J.~M.~Cline, E.~Filotas, J.~Matias, G.~D.~Moore,
  JHEP {\bf 0203}, 043 (2002).
  [hep-ph/0201082].


\bibitem{k2k}{\tt http://neutrino.kek.jp/}

\bibitem{minos}{\tt http://www-numi.fnal.gov/}

\bibitem{opera}{\tt http://operaweb.lngs.infn.it/?lang=en}

\bibitem{Giunti:2011hn}
  C.~Giunti, M.~Laveder,
  [arXiv:1109.4033 [hep-ph]].

\bibitem{Altschul:2006uw}
  B.~Altschul,
  Astropart.\ Phys.\  {\bf 28}, 380-384 (2007).
  [hep-ph/0610324].

\end{thebibliography}
\end{document}